\PassOptionsToPackage{table}{xcolor}

\documentclass[acmsmall,screen,nonacm]{acmart}

\usepackage{threeparttable}
\usepackage{tabularx}
\usepackage{multirow}
\usepackage{placeins}

\hypersetup{hidelinks}
\usepackage{fontawesome5}
\usepackage[most]{tcolorbox}

\newcolumntype{L}[1]{>{\raggedright\arraybackslash}p{#1}}
\newcolumntype{P}[1]{>{\centering\arraybackslash}p{#1}}
\newcolumntype{Y}{>{\centering\arraybackslash}X}

\newcommand{\rqanswericon}{\faLightbulb}

\newtcolorbox{rqanswerbox}[1][]{
    enhanced,
    breakable,
    colback=white,
    colframe=white,
    boxrule=0pt,
    arc=0pt,
    left=8pt,
    right=4pt,
    top=4pt,
    bottom=4pt,
    borderline west={1.5pt}{0pt}{black},
    #1
}

\setcopyright{none}
\renewcommand\footnotetextcopyrightpermission[1]{}

\begin{document}

\title{Memory Matters, but Retrieval Is Not Enough:
An Empirical Study of Memory Systems}

\author{Muhammad Waseem}
\email{muhammad.waseem@tuni.fi}
\affiliation{%
  \institution{Tampere University}
  \city{Tampere}
  \country{Finland}
}

\author{Md Nasir Uddin Shuvo}
\email{nasir.shuvo@tuni.fi}
\affiliation{%
  \institution{Tampere University}
  \city{Tampere}
  \country{Finland}
}

\author{Md Aidul Islam}
\email{mdaidul.islam@tuni.fi}
\affiliation{%
  \institution{Tampere University}
  \city{Tampere}
  \country{Finland}
}

\author{Sampo Autio}
\email{sampo.autio@etteplan.com}
\affiliation{%
  \institution{Etteplan}
  \city{Espoo}
  \country{Finland}
}

\author{Juho Poteri}
\email{juho.poteri@etteplan.com}
\affiliation{%
  \institution{Etteplan}
  \city{Espoo}
  \country{Finland}
}

\author{Arto Salminen}
\email{arto.salminen@etteplan.com}
\affiliation{%
  \institution{Etteplan}
  \city{Espoo}
  \country{Finland}
}

\author{Pekka Abrahamsson}
\email{pekka.abrahamsson@tuni.fi}
\affiliation{%
  \institution{Tampere University}
  \city{Tampere}
  \country{Finland}
}

\begin{abstract}
Large language model (LLM)-based systems increasingly use external memory to retain and reuse information across iterative interactions. This capability is relevant to AI-supported software evolution and engineering workflows, in which later steps depend on earlier decisions, constraints, and intermediate results. However, evidence remains limited about whether implemented memory systems improve performance on interdependent tasks or mainly increase retrieval activity. We conduct a matched comparison of a stateless baseline and three memory systems, Mem0, Hindsight, and Acontext, across 40 MemoryArena-based interdependent tasks, three LLMs, and three repeated executions, yielding 1,440 task-level observations. We examine partial progress, strict completion, Pass@$k$, position-specific and LLM-dependent effects, and observable retrieval and application behavior. Mem0 and Hindsight increase mean progress from 0.238 for the stateless baseline to 0.316 and 0.311, respectively, whereas Acontext provides a smaller gain that does not remain significant under the task-level analysis. Strict completion remains near the floor. Gains emerge at specific subtask positions and vary across LLMs, with the largest gains for GPT-5.5 and no reliable gain for Gemini 3.5 Flash. Retrieval volume alone does not explain performance: Acontext retrieves the most records but has the lowest answer overlap and marked usage, while Mem0 retrieves fewer records and, conditional on successful retrieval, has the highest values on both diagnostics. These findings show that memory can support partial progress without ensuring reliable task completion and should be evaluated as part of the complete LLM--memory configuration.

\end{abstract}

\maketitle

\section{Introduction}
\label{sec:introduction}

Engineering LLM-based systems for multi-step work requires maintaining information established during earlier interactions. Such systems increasingly combine reasoning, tool use, feedback-driven refinement, and interaction with external environments~\cite{yao2023react,schick2023toolformer,shinn2023reflexion,peng2023llmagentsurvey,luo2025llmagentsurvey}. This requirement is particularly relevant to software engineering workflows. Change impact analysis, requirements revision, refactoring, architecture decision maintenance, and technical documentation involve sequences in which later actions depend on earlier decisions, constraints, feedback, tool outputs, and intermediate results. Losing or misapplying this information can cause later outputs to overlook relevant constraints or conflict with prior decisions. Maintaining continuity across dependent steps is therefore important for AI-supported software engineering workflows.

External memory offers one approach to maintaining such continuity, but its effectiveness involves more than retaining additional context. A memory-enabled system must determine what information to store, how to represent and update it, what to retrieve for a later step, and how the retrieved information should influence the current response~\cite{packer2023memgpt,he2026memoryarena,zhang2025memorysurvey}. Failures at any stage can limit the value of memory. A system may fail to retrieve an important constraint, return outdated or weakly related information, or retrieve useful information without applying it effectively. Memory must therefore be evaluated as an end-to-end systems capability whose performance depends on the interaction between the memory system, the LLM, and the evaluation workload.

\noindent\textbf{Industrial Motivation.}
This study is motivated by an academic and industrial collaboration with Etteplan\footnote{\url{https://www.etteplan.com/}}, a technology service company working across engineering, software, embedded systems, and technical communication. Etteplan is investing in the development and adoption of LLM-based technologies for software-intensive and engineering projects. This investment raises practical questions about which memory system should be combined with a given LLM, whether the expected benefits persist across configurations, and which observable indicators can reveal ineffective retrieval or application. Before memory-enabled systems can be incorporated into workflows in which decisions, constraints, feedback, tool outputs, and domain knowledge evolve over time, their benefits and failure patterns must be systematically evaluated under common conditions. This industrial need motivates the evaluation presented in this paper, while the empirical study itself uses benchmark tasks rather than proprietary company data or a production deployment.

Research has produced several approaches to memory in LLM-based systems, including external context management~\cite{packer2023memgpt}, episodic reflection~\cite{shinn2023reflexion}, retrieval-augmented generation~\cite{lewis2020rag}, and structured or learned memory management~\cite{xu2025amem,wang2025mirix,yan2026memoryr1}. Memory benchmarks have also progressed from conversational recall toward interactive and interdependent tasks in which information obtained earlier must support later decisions~\cite{maharana2024locomo,wu2025longmemeval,hu2026memoryagentbench,he2026memoryarena}. However, systematic evidence about implemented memory systems under matched evaluation conditions remains limited. Existing evaluations often differ in their tasks, LLMs, memory interfaces, execution pipelines, and scoring procedures. These differences make it difficult to determine whether observed variation is associated with the memory system, the LLM, or the surrounding evaluation configuration. Aggregate task scores also provide limited insight into where memory-associated gains emerge within a dependent task or how retrieval and application behavior relate to the observed outcomes.

We address this gap through an empirical study built around a matched comparative evaluation of three implemented memory systems and a stateless baseline. We evaluate Mem0, Hindsight, Acontext, and the stateless condition on 40 interdependent tasks from the bundled-shopping domain of MemoryArena using three LLMs: GPT-5.5, Claude Opus 4.8, and Gemini 3.5 Flash. Each combination of memory condition, LLM, and task is executed three times, producing 1,440 task-level observations. We analyze the evaluated configurations at four complementary levels: overall task performance, progression across six dependent subtasks, variation across LLMs, and observable retrieval and application patterns. The explicit dependencies between MemoryArena task steps provide a common benchmark setting for examining whether information from earlier interactions supports later decisions. We use these tasks to study cross-step continuity rather than treating them as direct proxies for specific software engineering activities.

Our findings show measurable but bounded advantages for memory-enabled configurations. The evaluated memory systems achieve higher partial task progress than the stateless baseline to different degrees, while strict end-to-end completion remains rare. Because successful completion requires satisfying all six dependent subtasks, subtask progression provides a more discriminating view of where configurations advance and where they fail. The observed differences also vary across LLMs and across positions within the dependent task sequence. Moreover, retrieval activity alone is not a sufficient indicator of effective memory behavior. Systems that retrieve more records do not necessarily achieve stronger task progress or higher measured application of retrieved information. These results support evaluating memory at the level of the complete LLM and memory-system configuration rather than treating it as a uniformly beneficial capability.

\noindent\textbf{Contributions.}
This paper makes the following contributions:

\begin{itemize}

\item An empirical comparison of three memory systems (Mem0, Hindsight, and Acontext) against a stateless baseline across three LLMs, comprising 1,440 task-level observations on 40 interdependent tasks.

\item Evidence that memory improves partial progress but not full-task completion, and that its gains depend on the subtask position and the selected LLM.

\item Evidence that retrieval volume does not indicate effective memory use, based on answer overlap, marked usage, and failure patterns, together with implications for evaluating and monitoring memory-enabled LLM systems.

\item A replication package containing the experiment code ~\cite{memory-replication} memory-replication and study dataset~\cite{memory-dataset}.

\end{itemize}
\section{Background and Related Work}
\label{sec:background}

Research on memory in LLM-based systems spans long-term conversational memory~\cite{zhong2024memorybank}, external context management~\cite{packer2023memgpt}, retrieval augmentation~\cite{lewis2020rag}, reflection over prior experience~\cite{shinn2023reflexion}, and structured or learned memory management~\cite{xu2025amem,wang2025mirix,yan2026memoryr1}. Across these directions, memory has evolved from the passive retention of interaction history into an active systems capability that determines what information is preserved, retrieved, updated, and made available for subsequent decisions.

\subsection{Memory Systems and Memory Management}
\label{sec:memory-systems}

Memory systems differ in the information they preserve and in how that information is made available during later interactions. MemoryBank stores, recalls, updates, and selectively forgets user-related information to support sustained interaction with LLM-based assistants~\cite{zhong2024memorybank}. MemGPT treats memory as virtual context management, moving information between the limited context window and external storage~\cite{packer2023memgpt}. Reflexion records verbal feedback and self-reflection from previous attempts so that prior experience can inform subsequent decisions~\cite{shinn2023reflexion}. Retrieval-augmented generation provides a broader foundation for retrieving external information during generation, although its traditional formulation does not necessarily maintain evolving memories from prior interactions~\cite{lewis2020rag}.

Implemented memory systems adopt different strategies for constructing and retrieving such information. Mem0 dynamically extracts, consolidates, and retrieves salient information from prior interactions~\cite{chhikara2025mem0}. Hindsight organizes long-term memory through retain, recall, and reflect operations~\cite{latimer2025hindsight}. Acontext represents learned information as reusable skill descriptions derived from stored agent sessions~\cite{acontext2026}. These systems form the memory-enabled conditions evaluated in this study.

Recent research has also introduced increasingly structured and adaptive memory management. A-MEM represents memories as interconnected notes and updates their relationships as new experiences accumulate~\cite{xu2025amem}. MIRIX organizes six memory types, including episodic, semantic, and procedural memory~\cite{wang2025mirix}. Memory-R1 formulates memory management as a learned decision process that determines when memory entries should be added, updated, deleted, or ignored~\cite{yan2026memoryr1}.


\subsection{Benchmarks and Evaluation of Memory}
\label{sec:memory-benchmarks}

Several benchmarks evaluate whether LLM-based systems can retain and recover information over extended interactions. LoCoMo evaluates long-term conversational memory through question answering, event summarization, and multimodal dialogue generation over multi-session conversations~\cite{maharana2024locomo}. LongMemEval evaluates information extraction, multi-session reasoning, temporal reasoning, knowledge updates, and abstention~\cite{wu2025longmemeval}. MemBench considers factual and reflective memory across participation and observation scenarios, together with effectiveness, efficiency, and capacity~\cite{tan2025membench}. MemoryAgentBench evaluates accurate retrieval, test-time learning, long-range understanding, and selective forgetting through incremental multi-turn interactions~\cite{hu2026memoryagentbench}. These benchmarks broaden memory evaluation beyond isolated factual recall by examining how information is maintained, updated, and selectively recovered over time.

Other benchmarks connect memory more directly with actions and dependent task execution. The LTM benchmark evaluates continual interaction through an extended conversation in which multiple tasks are interleaved~\cite{castillo2024beyondprompts}. MemoryArena introduces multi-session tasks composed of explicitly interdependent subtasks, where information obtained during earlier interactions is required for later decisions~\cite{he2026memoryarena}. Its results show that strong performance on conversational-memory benchmarks does not necessarily transfer to agentic settings in which remembered information must support subsequent actions.

System-level evaluations further indicate that memory performance depends on the complete configuration and evaluated workload. MemoryLake compares structured memory, Mem0, vector retrieval, and long-context configurations across MemoryArena domains under a shared LLM and execution protocol~\cite{zhan2026memorylake}. Its results show that the relative performance of memory approaches varies across task types, supporting a workload-dependent view of memory performance. The evaluation uses one LLM and reports primarily domain-level process and success measures. Evidence therefore remains limited regarding whether differences among implemented memory systems persist across LLMs, where they emerge within dependent task sequences, and how observable retrieval and application patterns relate to performance.

\subsection{Positioning of This Study}
\label{sec:related-positioning}

This study differs from prior studies in three respects. First, it compares implemented memory systems under a single matched protocol across three LLMs, with paired significance tests and effect sizes, whereas existing system-level comparisons rely on one LLM and report point estimates~\cite{zhan2026memorylake}. Second, it locates memory-associated differences at specific positions within dependent task sequences rather than reporting only aggregate outcomes. Third, it relates task outcomes to observable retrieval and application behavior, distinguishing retrieval volume from answer overlap and marked usage. This design exposes effects that single-LLM aggregate evaluations cannot reveal, including memory gains that depend on the selected LLM and retrieval activity that does not correspond to effective application.
\section{Study Design}
\label{sec:study-design}

\begin{figure*}[t]
\centering
\includegraphics[width=\textwidth]{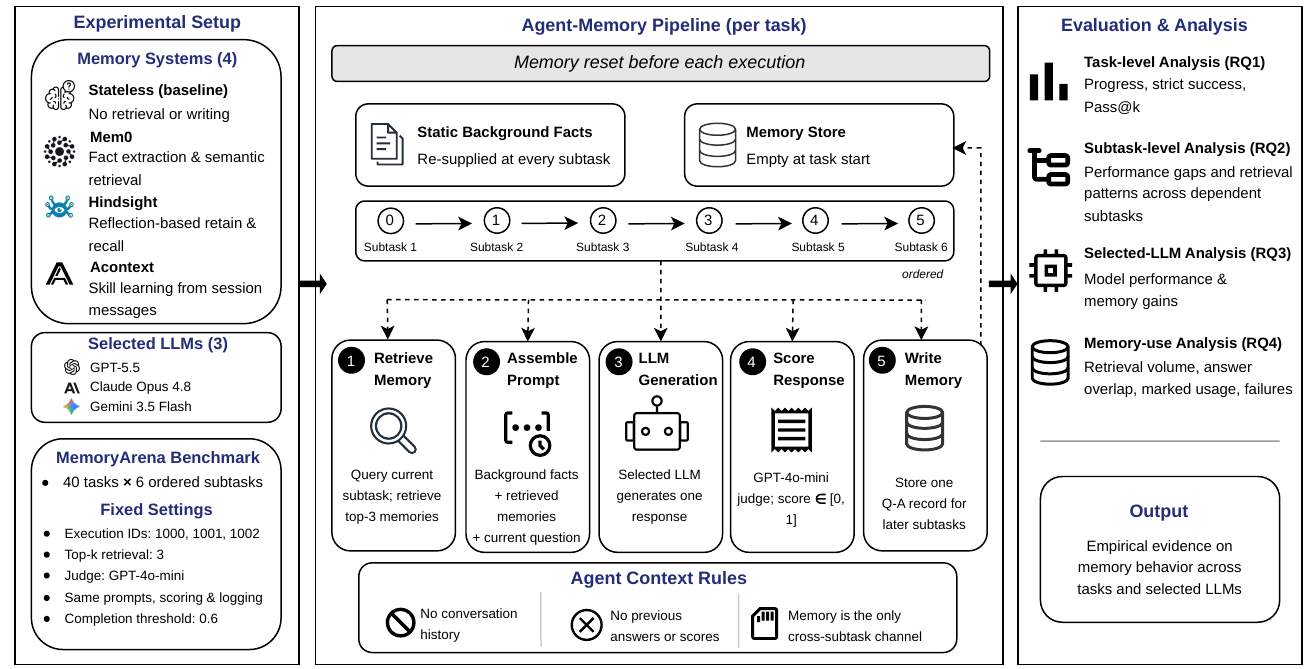}
\caption{Matched comparative evaluation of three implemented memory systems and a stateless baseline across three LLMs.}
\label{fig:study-design}
\end{figure*}

We conduct a matched comparative evaluation of three implemented memory systems and a stateless baseline, varying the memory condition and response-generating LLM. We hold the benchmark tasks, prompt structure, subtask sequence, retrieval limit, record presentation, write timing, and scoring procedure constant, while each memory system retains its native representation, retrieval procedure, and internal processing. Comparisons therefore concern complete LLM and memory-system configurations rather than isolated mechanisms. Figure~\ref{fig:study-design} summarizes the study design.

\subsection{Research Questions}
\label{sec:research-questions}

The study addresses four research questions:

\begin{itemize}
\item \textbf{RQ1:} How does task performance differ between memory-enabled configurations and the stateless baseline?
\item \textbf{RQ2:} How do task performance and retrieval patterns vary across dependent subtask positions?
\item \textbf{RQ3:} How do memory-associated performance differences vary across LLMs?
\item \textbf{RQ4:} What retrieval-volume, answer-overlap, non-contradiction, marked-usage, and failure patterns characterize the evaluated memory systems?
\end{itemize}

RQ1 assesses overall task performance; RQ2 examines performance and retrieval across six dependent positions; RQ3 examines variation across LLM and memory-system combinations; and RQ4 characterizes operational memory diagnostics. Because the logs do not preserve complete record lineage, RQ2 and RQ4 support configuration-level comparisons rather than causal claims about individual memories.

\subsection{Benchmark and Task Setting}
\label{sec:benchmark-task-setting}

We use the \texttt{bundled\_shopping} domain of MemoryArena~\cite{he2026memoryarena}, which requires selecting related products across multi-session tasks. From 150 labeled tasks, we select 40 using a fixed pseudorandom shuffle with seed 42 and use them across all configurations. Each task contains six ordered subtasks in which later decisions may depend on earlier attributes, constraints, prices, and preferences. We use the MemoryArena dataset but not its official runner. Our harness controls prompt construction, LLM and memory-system invocation, logging, and scoring. Results are therefore MemoryArena-based rather than official benchmark scores. Each task, execution, memory condition, and LLM combination receives an isolated namespace, so memory is not shared across tasks or executions. The study evaluates within-task continuity rather than cross-task transfer or persistent user memory.

\subsection{LLM and Memory Execution Pipeline}
\label{sec:llm-memory-execution-pipeline}

The evaluated LLM receives one subtask at a time. For each subtask, the pipeline performs five steps:

\begin{itemize}
\item \textbf{Retrieve:} The current question is submitted as the retrieval query. A memory-enabled system may return up to three records, whereas the stateless baseline returns none.
\item \textbf{Construct the prompt:} The harness combines the static task background, retrieved records, and current question using a common prompt structure.
\item \textbf{Generate:} The evaluated LLM generates one response.
\item \textbf{Evaluate:} The response is scored using the procedure described in Section~\ref{sec:response-scoring}.
\item \textbf{Write:} For memory-enabled conditions, the current question and generated answer are submitted to the memory system.
\end{itemize}

The evaluated LLM does not receive the preceding conversation, earlier answers, or judge scores. Static background information is repeated at every subtask, while information introduced during previous subtasks can be carried forward only through the memory-system interface. Judge outputs are neither shown to the evaluated LLM nor written to memory. The findings should therefore be interpreted under this fixed write-and-retrieve protocol.

\subsection{Experimental Conditions}
\label{sec:experimental-conditions}

We evaluate one stateless condition and three memory-enabled systems:

\begin{itemize}
\item \textbf{Stateless} performs no memory writing or retrieval.
\item \textbf{Mem0} extracts information from submitted records and supports semantic retrieval~\cite{chhikara2025mem0}.
\item \textbf{Hindsight} provides a reflection-oriented retain-and-recall mechanism~\cite{latimer2025hindsight}.
\item \textbf{Acontext} derives skill-like descriptions from stored session messages and retrieves the resulting descriptions~\cite{acontext2026}.
\end{itemize}

The systems are integrated through a common adapter supporting initialization, namespace management, retrieval, and writing. All memory-enabled systems use GPT-4o-mini for their internal LLM-based processing and \texttt{text-embedding-3-small} for embeddings. Acontext initializes each namespace with two built-in, server-supplied skill descriptions before the first benchmark interaction. These records are treated as part of the evaluated Acontext configuration.

We evaluate every condition with GPT-5.5, Claude Opus 4.8, and Gemini 3.5 Flash. Exact model snapshots, dependency versions, prompts, and inference settings are provided in the replication package.

Each combination of memory condition, LLM, and task is executed three times. The execution identifiers 1000, 1001, and 1002 define separate namespaces but are not passed as sampling seeds to the LLM APIs. These executions therefore represent repeated stochastic observations rather than deterministically reproducible model generations. The evaluation contains \(4 \times 3 \times 40 \times 3 = 1{,}440\) task-level observations and \(8{,}640\) subtask evaluations.

\subsection{Measures}
\label{sec:evaluation-measures}

\subsubsection{Response Scoring}
\label{sec:response-scoring}

All responses are scored by GPT-4o-mini using temperature 0. For each subtask, the judge receives the question, expected answer, evaluation rubric, and generated response and returns a score between 0 and 1. Unparseable responses receive a score of zero. A subtask is classified as completed when its score is at least 0.6. This cutoff is fixed by the evaluation harness and is not estimated from the present observations. We treat the resulting scores as operational evaluation measures rather than human-validated ground truth.

\subsubsection{Task Performance}
\label{sec:task-performance-measures}

For each task-level observation, we calculate:

\begin{itemize}
\item \textit{Score}: the mean judge score across the six subtasks.
\item \textit{\texttt{progress\_score}}: the proportion of subtasks with a score of at least 0.6.
\item \textit{Strict success}: an indicator equal to one when all six subtasks meet the threshold.
\end{itemize}

The primary performance outcome for RQ1 and RQ3 is \texttt{progress\_score}. We also report Pass@\(k\) for \(k \in \{1,2,3\}\). For an LLM-task cell with \(c\) strict successes among three repeated executions, the Pass@\(k\) contribution is \(1-\binom{3-c}{k}/\binom{3}{k}\), averaged over the 120 LLM-task cells in each memory condition.

\subsubsection{Within-Task Progression}
\label{sec:within-task-progression}

At each subtask position, we examine the judge score, thresholded completion, and number of retrieved records. Because the event logs do not preserve repeated-execution identifiers, RQ2 averages the three pooled observations for each memory-condition, LLM, task, and position combination. We use the continuous judge score for position-level inference because \texttt{progress\_score} summarizes completion across the complete six-subtask task. The analysis therefore compares subtask positions but does not reconstruct exact within-execution trajectories.

\subsubsection{Memory-Behavior Diagnostics}
\label{sec:memory-behavior-diagnostics}

For memory-enabled observations, we analyze retrieval and write counts together with four operational indicators:

\begin{itemize}
\item \textit{Answer overlap}: the mean token-set Jaccard overlap between retrieved records and the generated response, with an LLM-based fallback assessment when the overlap is zero;
\item \textit{Non-contradiction}: a heuristic indicator of whether retrieved records conflict with an offline task fact ledger that is not shown to the evaluated LLM or memory system;
\item \textit{Marked usage}: the proportion of retrieved identifiers reported as used when \texttt{used\_memory\_ids} are available, otherwise estimated using an answer-overlap threshold and an LLM-based fallback; and
\item \textit{Failure category}: an ordered classification of retrieval, writing, contradiction, usage, consistency, environment, or residual reasoning failures.
\end{itemize}

These indicators correspond to the logged relevance, correctness, usage, and failure fields but use names that more accurately reflect their operational definitions. They are diagnostic measures rather than independently human-validated assessments of memory quality.

\subsection{Data Analysis}
\label{sec:data-analysis}

For RQ1, we average the three repeated executions within each memory-condition, LLM, and task cell and then average across LLMs. This produces 40 paired task means for each comparison with the stateless baseline. We apply two-sided paired \(t\)-tests to \texttt{progress\_score}, use Holm correction across the three comparisons, and report paired \(d_z\). Strict success and Pass@\(k\) are reported descriptively.

For RQ2, we average the three pooled event scores for each memory-condition, LLM, task, and subtask position. At each position, this produces 120 matched LLM-task observations per comparison. We apply paired \(t\)-tests to the judge score, report \(d_z\), and use Holm correction across the 18 memory-system and position comparisons. Thresholded completion and retrieval volume are analyzed descriptively across the six positions.

For RQ3, we average repeated executions within each task, memory condition, and LLM cell. We apply a task-blocked repeated-measures ANOVA with memory condition and LLM as within-task factors, report Greenhouse--Geisser-adjusted \(p\)-values and partial \(\eta_p^2\), and conduct nine Holm-adjusted paired comparisons with the corresponding stateless conditions, reporting \(d_z\).

For RQ4, we report descriptive means over valid observations. For each pairwise comparison of answer overlap and marked usage, we first compute differences within matched LLM-task-execution cells having valid values for both systems and then average those differences within each task. This produces 40 task-level differences per contrast. We apply paired \(t\)-tests, use Holm correction across the six contrasts, and report task-level \(d_z\). Associations with task progress use Pearson correlations on complete-case observations after subtracting the corresponding task mean from both variables. Significance is assessed using the residual degrees of freedom from the corresponding task-fixed-effects model, with Holm correction across the 11 estimable associations. Acontext non-contradiction is constant, so its correlation is undefined. Failure-category frequencies are reported descriptively. Missing diagnostic values caused by empty retrieval are excluded rather than converted to zero.

Incomplete event traces were replaced by complete reruns under the same protocol, while task-level analyses retain the original observations. All 120 Hindsight and GPT-5.5 observations invoked the adapter's local token-overlap fallback at least once. Because its trigger was not logged, we retain this configuration and repeat the relevant analyses without it. The replication package documents execution, substitution, and version details. All tests are two-sided with \(\alpha=0.05\).
\section{Results}
\label{sec:results}

This section reports the empirical results for the four research questions.

\subsection{Task-Level Performance of Memory Systems (RQ1)}
\label{sec:rq1-results}

\begin{figure*}[!t]
\centering
\begin{minipage}[t]{0.50\textwidth}
\centering
\includegraphics[width=\linewidth]{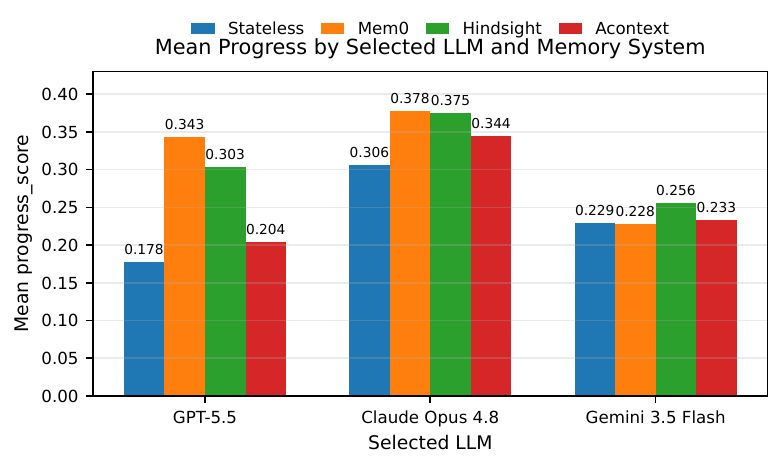}
\\[-0.3em]
\textbf{(a) Mean progress by LLM and memory system}
\end{minipage}
\hfill
\begin{minipage}[t]{0.48\textwidth}
\centering
\includegraphics[width=\linewidth]{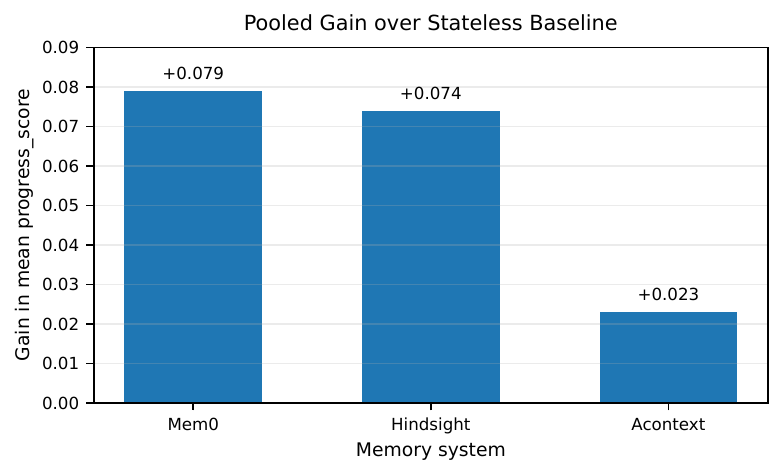}
\\[-0.3em]
\textbf{(b) Pooled gain over stateless}
\end{minipage}

\caption{Task-level performance across memory conditions.
Panel~(a) reports the mean \textit{progress\_score} of each memory condition for each LLM.
Panel~(b) reports the pooled gain in mean \textit{progress\_score} of each memory-enabled system relative to the stateless baseline across all LLMs.}
\label{fig:rq1-memory-effect}
\end{figure*}

To answer RQ1, we compare the task-level performance of the stateless baseline and the three memory-enabled systems across all LLMs and tasks.  Table~\ref{tab:memory-performance} summarises overall performance for each memory condition, whereas Figure~\ref{fig:rq1-memory-effect} shows performance by LLM and the pooled gain over the stateless baseline.

\begin{table*}[!t]
\centering
\caption{Performance of memory systems relative to the stateless baseline.}
\label{tab:memory-performance}
\scriptsize
\begin{threeparttable}
\renewcommand{\arraystretch}{1.15}
\setlength{\tabcolsep}{4pt}
\begin{tabularx}{\textwidth}{lYYYYYYYY}
\toprule
\rowcolor{gray!15}
\textbf{Memory system} &
\textbf{Mean progress} &
\textbf{Median progress} &
\textbf{IQR} &
\textbf{Pass@1} &
\textbf{Pass@2} &
\textbf{Pass@3} &
\textbf{Gain vs. stateless} &
\textbf{Holm $p$ vs. stateless} \\
\midrule
Stateless  & 0.238 & 0.167 & 0.167 & 0.0\% & 0.0\% & 0.0\% & 0.000 & N/A \\
Mem0       & \textbf{0.316} & \textbf{0.333} & \textbf{0.333} & 1.4\% & 1.9\% & 2.5\% & +0.079 & $<0.001$ \\
Hindsight  & 0.311 & \textbf{0.333} & \textbf{0.333} & 0.8\% & 1.7\% & 2.5\% & +0.074 & $<0.001$ \\
Acontext   & 0.261 & 0.167 & 0.167 & 0.0\% & 0.0\% & 0.0\% & +0.023 & $0.0125$ \\
\bottomrule
\end{tabularx}

\begin{tablenotes}
\scriptsize
\item Note. Each memory condition contains 360 task-level observations: 3 LLMs $\times$ 40 tasks $\times$ 3 repeated executions. Mean, median, and IQR are computed over these observations. Pass@$k$ is computed within each of the 120 LLM--task cells containing three repeated executions and then averaged across the cells. Gain and Holm-adjusted $p$-values are based on 40 paired task means. The three executions are first averaged within each memory-condition, LLM, and task combination, after which the three LLM means are averaged within each task and memory condition. N/A indicates that no comparison applies to the baseline.
\end{tablenotes}
\end{threeparttable}
\end{table*}

{\faChartLine} {\textbf{Memory-enabled configurations achieve higher partial task progress}:}
Mem0 achieves the highest mean \textit{progress\_score} at 0.316, followed closely by Hindsight at 0.311, compared with 0.238 for the stateless baseline. Their respective gains of +0.079 and +0.074 are statistically significant with medium-to-large paired effect sizes (Mem0: $d_z=0.79$, Holm $p<0.001$; Hindsight: $d_z=0.81$, Holm $p<0.001$). Acontext produces a smaller gain of +0.023, but the difference remains statistically significant under the task-aware analysis ($d_z=0.41$, Holm $p=0.0125$).

{\faExclamationTriangle} {\textbf{Full-task success remains rare}:}
Higher partial progress does not translate into reliable end-to-end completion. Pass@1 remains below 1.5\% for every memory-enabled system, while Pass@3 reaches only 2.5\% for Mem0 and Hindsight. Neither the stateless baseline nor Acontext completes any task across the three repeated executions.

{\faLayerGroup} {\textbf{Performance differences vary across LLMs}:}
Figure~\ref{fig:rq1-memory-effect} shows that the pooled results conceal substantial variation across LLM and memory-system combinations. We examine this variation directly in RQ3.

\begin{rqanswerbox}
\noindent\rqanswericon\hspace{0.35em}\textbf{Answer to RQ1:}
Memory-enabled configurations achieve higher partial task progress than the stateless baseline, but they do not provide reliable full-task completion. Mem0 and Hindsight produce the largest gains, while Acontext provides a smaller but statistically significant gain. Pass@3 reaches only 2.5\%, showing that memory helps configurations advance through dependent tasks but remains insufficient for dependable end-to-end success.
\end{rqanswerbox}

\subsection{Performance and Retrieval Across Dependent Subtasks (RQ2)}
\label{sec:rq2-results}

\begin{figure}[tbp]
\centering
\begin{minipage}[t]{0.49\linewidth}
    \centering
    \includegraphics[width=\linewidth]{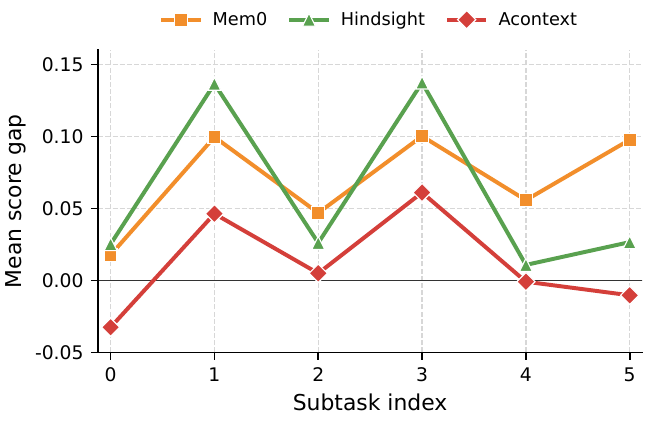}
    \par\smallskip
    \textbf{(a) Memory-versus-stateless gap}
\end{minipage}\hfill
\begin{minipage}[t]{0.49\linewidth}
    \centering
    \includegraphics[width=\linewidth]{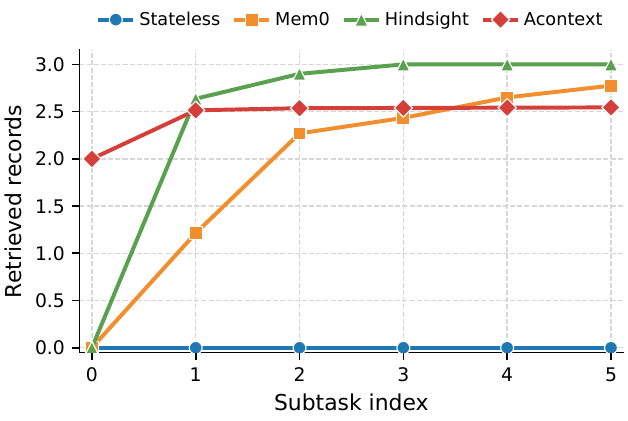}
    \par\smallskip
    \textbf{(b) Retrieved records}
\end{minipage}
\caption{Performance and retrieval across six ordered subtasks. Panel~(a) reports mean memory-versus-stateless score gaps, and Panel~(b) reports the average number of retrieved records.}
\label{fig:rq2-memory-reuse}
\end{figure}

To answer RQ2, we examine how performance and retrieval behavior vary across the six dependent subtask positions. Table~\ref{tab:rq2-paired-contrasts} reports paired memory-versus-stateless contrasts at each position. Figure~\ref{fig:rq2-memory-reuse} presents the corresponding performance gaps and retrieval patterns.

\begin{table*}[!t]
\centering
\caption{Paired memory-versus-stateless contrasts across dependent subtasks.}
\label{tab:rq2-paired-contrasts}
\scriptsize
\begin{threeparttable}
\renewcommand{\arraystretch}{1.12}
\setlength{\tabcolsep}{3pt}

\begin{tabular}{
L{0.17\textwidth}
P{0.075\textwidth}
P{0.055\textwidth}
P{0.095\textwidth}
P{0.095\textwidth}
P{0.075\textwidth}
P{0.095\textwidth}
P{0.095\textwidth}
P{0.075\textwidth}
}
\toprule
\rowcolor{gray!15}
\textbf{Contrast}
& \textbf{Subtask index}
& \textbf{Pairs ($n$)}
& \textbf{Mean gap ($\Delta$ score)}
& \textbf{SD of gap}
& \textbf{$t$-statistic}
& \textbf{$p$-value}
& \textbf{Holm $p$-value}
& \textbf{Effect size ($d_z$)} \\
\midrule

\multirow[c]{6}{0.17\textwidth}{\raggedright Mem0 vs stateless}
& 0 & 120 & +0.018 & 0.233 & 0.84 & 0.404 & 1.000 & 0.08 \\
& 1 & 120 & \textbf{+0.100} & 0.325 & 3.37 & 0.001 & \textbf{0.016} & 0.31 \\
& 2 & 120 & +0.047 & 0.364 & 1.41 & 0.161 & 1.000 & 0.13 \\
& 3 & 120 & \textbf{+0.100} & 0.354 & 3.11 & 0.002 & \textbf{0.033} & 0.28 \\
& 4 & 120 & +0.056 & 0.335 & 1.82 & 0.071 & 0.778 & 0.17 \\
& 5 & 120 & \textbf{+0.098} & 0.319 & 3.36 & 0.001 & \textbf{0.016} & 0.31 \\
\midrule

\multirow[c]{6}{0.17\textwidth}{\raggedright Hindsight vs stateless}
& 0 & 120 & +0.025 & 0.198 & 1.40 & 0.164 & 1.000 & 0.13 \\
& 1 & 120 & \textbf{+0.137} & 0.392 & 3.82 & $<0.001$ & \textbf{0.004} & 0.35 \\
& 2 & 120 & +0.026 & 0.272 & 1.05 & 0.294 & 1.000 & 0.10 \\
& 3 & 120 & \textbf{+0.138} & 0.359 & 4.19 & $<0.001$ & \textbf{$<0.001$} & 0.38 \\
& 4 & 120 & +0.011 & 0.252 & 0.47 & 0.639 & 1.000 & 0.04 \\
& 5 & 120 & +0.027 & 0.341 & 0.86 & 0.393 & 1.000 & 0.08 \\
\midrule

\multirow[c]{6}{0.17\textwidth}{\raggedright Acontext vs stateless}
& 0 & 120 & -0.033 & 0.288 & -1.24 & 0.219 & 1.000 & -0.11 \\
& 1 & 120 & +0.046 & 0.271 & 1.87 & 0.063 & 0.762 & 0.17 \\
& 2 & 120 & +0.005 & 0.236 & 0.23 & 0.817 & 1.000 & 0.02 \\
& 3 & 120 & +0.061 & 0.303 & 2.21 & 0.029 & 0.378 & 0.20 \\
& 4 & 120 & -0.001 & 0.290 & -0.03 & 0.975 & 1.000 & 0.00 \\
& 5 & 120 & -0.010 & 0.231 & -0.49 & 0.627 & 1.000 & -0.04 \\
\bottomrule
\end{tabular}

\begin{tablenotes}
\scriptsize
\item Note. $\Delta$ score is the paired difference between a memory-enabled condition and stateless within the same LLM, task, and subtask position. The three pooled scores are averaged within each cell, producing 120 matched LLM--task observations per position. Execution-specific trajectories cannot be reconstructed because the event logs do not preserve execution identifiers. Holm correction is applied jointly across the 18 contrasts. Subtasks are indexed from 0 to 5. Bold values remain significant after correction.
\end{tablenotes}

\end{threeparttable}
\end{table*}

{\faChartLine} {\textbf{Performance gains emerge at specific positions}:}
Mem0 achieves significant gains at subtasks 1, 3, and 5, ranging from +0.098 to +0.100, with effect sizes between $d_z=0.28$ and $d_z=0.31$. Hindsight achieves its largest gains at subtasks 1 and 3 ($\Delta=+0.137$ and $+0.138$). Neither system differs reliably from stateless at subtask 0. A task-level sensitivity analysis retains the subtask-1 and subtask-3 gains for both systems, while the Mem0 subtask-5 result becomes borderline (Holm $p=0.056$).

{\faLayerGroup} {\textbf{Retrieval volume increases across the task}:}
Mem0 retrieves no records at subtask 0, an average of 1.214 at subtask 1, and 2.775 by subtask 5. Hindsight retrieves an average of 2.636 records at subtask 1 and reaches the limit of three records from subtask 3 onward. These patterns reflect the accumulation of within-task memory records.

{\faExclamationTriangle} {\textbf{More retrieval does not consistently correspond to higher performance}:}
Hindsight retrieves three records at subtasks 3, 4, and 5, but only its subtask-3 difference is significant among these positions. Mem0 also retrieves progressively more records, yet its differences are not significant at subtasks 2 and 4. The Hindsight gains at subtasks 1 and 3 also do not remain significant after excluding the fallback-affected Hindsight and GPT-5.5 condition.

{\faLayerGroup} {\textbf{Acontext shows no reliable position-specific gain}:}
Acontext retrieves two server-supplied records at subtask 0, before the harness writes any benchmark interaction. Its retrieval count subsequently remains stable, increasing from 2.514 at subtask 1 to 2.544 at subtask 5. Despite this sustained retrieval activity, none of its performance differences remains significant after Holm correction.

\begin{rqanswerbox}
\noindent\rqanswericon\hspace{0.35em}\textbf{Answer to RQ2:}
Memory-associated performance differences are position-specific. Mem0 achieves reliable gains at subtasks 1 and 3, with weaker evidence at subtask 5. Hindsight achieves gains at subtasks 1 and 3, although these results are sensitive to the fallback-affected GPT-5.5 condition. Acontext shows no reliable position-specific gain. Retrieval increases across later subtasks, but retrieval volume alone does not explain when memory improves performance.
\end{rqanswerbox}

\subsection{LLM-Specific Memory Effects (RQ3)}
\label{sec:rq3-results}

\begin{figure}[tbp]
\centering
\includegraphics[width=0.80\linewidth]{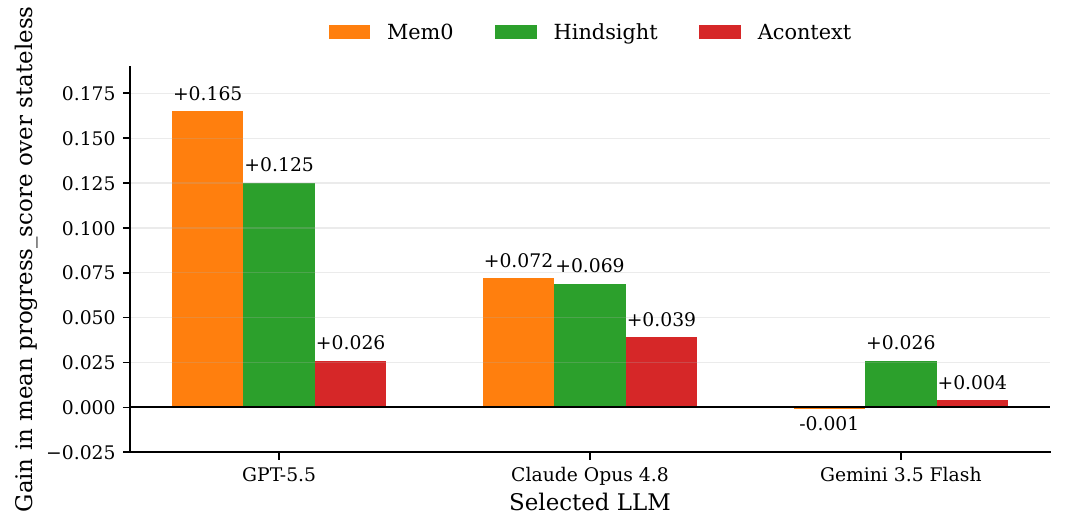}
\caption{LLM-specific memory gains over the stateless baseline. Bars report the difference in mean \textit{progress\_score} between each memory-enabled system and the stateless baseline within the same LLM.}
\label{fig:rq3-model-memory-progress}
\end{figure}

To answer RQ3, we examine whether the effects of the evaluated memory systems vary across the selected LLMs. Table~\ref{tab:rq3-llm-effects} reports overall LLM performance, memory-system-specific progress, and paired gains over the corresponding stateless baseline. Figure~\ref{fig:rq3-model-memory-progress} visualizes these LLM-specific gains.

\begin{table*}[!t]
\centering
\caption{LLM-specific effects of memory systems on task progress.}
\label{tab:rq3-llm-effects}
\scriptsize
\begin{threeparttable}
\renewcommand{\arraystretch}{1.12}
\setlength{\tabcolsep}{3pt}

\begin{tabular}{
L{0.13\textwidth}
P{0.09\textwidth}
P{0.06\textwidth}
P{0.07\textwidth}
P{0.10\textwidth}
P{0.065\textwidth}
P{0.10\textwidth}
P{0.065\textwidth}
P{0.10\textwidth}
P{0.065\textwidth}
}
\toprule
\rowcolor{gray!15}
\textbf{Selected LLM}
& \textbf{Overall mean (SD)}
& \textbf{Success rate}
& \textbf{Stateless mean}
& \textbf{Mem0 mean / gain}
& \textbf{Mem0 Holm $p$}
& \textbf{Hindsight mean / gain}
& \textbf{Hindsight Holm $p$}
& \textbf{Acontext mean / gain}
& \textbf{Acontext Holm $p$} \\
\midrule

GPT-5.5
& 0.257 (0.243)
& 1.3\%
& 0.178
& 0.343 / \textbf{+0.165}
& \textbf{$<0.001$}
& 0.303 / \textbf{+0.125}
& \textbf{0.003}
& 0.204 / +0.026
& 0.363 \\

Claude Opus 4.8
& 0.351 (0.219)
& 0.4\%
& 0.306
& 0.378 / \textbf{+0.072}
& \textbf{0.033}
& 0.375 / \textbf{+0.069}
& \textbf{0.04996}
& 0.344 / +0.039
& 0.363 \\

Gemini 3.5 Flash
& 0.236 (0.182)
& 0.0\%
& 0.229
& 0.228 / -0.001
& 1.000
& 0.256 / +0.026
& 0.802
& 0.233 / +0.004
& 1.000 \\

\bottomrule
\end{tabular}

\begin{tablenotes}
\scriptsize
\item Note. Overall means, SDs, and success rates are computed over 480 runs per LLM. Memory-system-specific means are computed over 120 runs per memory-system and LLM combination. Gain denotes the difference in mean \textit{progress\_score} from the stateless baseline within the same LLM. For inference, the three repeated executions are averaged within each task, producing 40 paired task-level observations per comparison. Holm correction is applied jointly across the nine memory-system-versus-stateless comparisons. Bold values indicate gains that remain significant after correction. The Claude Opus 4.8 and Hindsight comparison is borderline, with an unrounded adjusted $p$-value of 0.04996.
\end{tablenotes}

\end{threeparttable}
\end{table*}

{\faLayerGroup} {\textbf{Overall progress differs across LLMs}:}
Claude Opus 4.8 achieves the highest overall mean \textit{progress\_score} (0.351), followed by GPT-5.5 (0.257) and Gemini 3.5 Flash (0.236). The task-blocked repeated-measures analysis identifies an overall LLM effect, $F(2,78)=13.71$, $p_{\mathrm{GG}}<0.001$, with partial $\eta_p^2=0.260$.

{\faChartLine} {\textbf{GPT-5.5 exhibits the largest memory gains}:}
Mem0 increases mean \textit{progress\_score} from 0.178 to 0.343, corresponding to a gain of +0.165 ($p_{\mathrm{Holm}}<0.001$, $d_z=0.766$). Hindsight increases progress to 0.303, corresponding to a gain of +0.125 ($p_{\mathrm{Holm}}=0.003$, $d_z=0.620$). Acontext provides a smaller gain of +0.026 that does not remain significant. Because all Hindsight and GPT-5.5 observations invoked the adapter's local fallback at least once, the Hindsight result should be interpreted as evidence about the evaluated configuration rather than an isolated effect of its native implementation.

{\faLayerGroup} {\textbf{Claude Opus 4.8 has the highest overall progress}:}
Its stateless baseline is 0.306. Mem0 increases progress to 0.378, corresponding to a gain of +0.072 ($p_{\mathrm{Holm}}=0.033$, $d_z=0.473$). Hindsight reaches 0.375, corresponding to a gain of +0.069 ($p_{\mathrm{Holm}}=0.04996$, $d_z=0.440$). The Hindsight result lies at the significance threshold and should therefore be interpreted cautiously. Acontext's gain of +0.039 does not remain significant.

{\faExclamationTriangle} {\textbf{Memory provides no reliable gain for Gemini 3.5 Flash}:}
Its stateless mean is 0.229. Mem0 produces a near-zero negative difference of -0.001, while Hindsight and Acontext produce gains of +0.026 and +0.004, respectively. None of these differences remains significant after correction. Although Acontext shows a small statistically significant pooled gain in RQ1, none of its LLM-specific contrasts remains significant after correction across the nine RQ3 comparisons.

{\faChartLine} {\textbf{Memory effects vary across LLMs}:}
The task-blocked interaction between memory system and LLM is statistically significant, $F(6,234)=4.19$, $p_{\mathrm{GG}}=0.0018$, with partial $\eta_p^2=0.097$. The interaction remains significant when Hindsight is excluded but not when GPT-5.5 is excluded. This sensitivity pattern indicates that the interaction is driven primarily by the comparatively large gains under GPT-5.5 rather than solely by the fallback-affected Hindsight configuration.

\begin{rqanswerbox}
\noindent\rqanswericon\hspace{0.35em}\textbf{Answer to RQ3:}
Memory effects vary across LLMs. Mem0 provides the largest gain for GPT-5.5 and a smaller but significant gain for Claude Opus 4.8. Hindsight follows the same general pattern, although its GPT-5.5 result is affected by the adapter fallback and its Claude Opus 4.8 result is borderline. None of the evaluated memory systems reliably improves Gemini 3.5 Flash, and Acontext provides no significant gain for any LLM. Memory effectiveness should therefore be evaluated at the level of the complete LLM and memory-system configuration.
\end{rqanswerbox}

\subsection{Memory Retrieval and Application Behavior (RQ4)}
\label{sec:rq4-results}

To answer RQ4, we characterize observable retrieval and application behavior using automatically derived diagnostics. This analysis includes only memory-enabled systems because the stateless baseline neither writes nor retrieves external memory. Table~\ref{tab:rq4-memory-quality} reports retrieval volume, answer overlap, non-contradiction, marked usage, failure labels, and task progress. Table~\ref{tab:rq4-quality-contrasts} reports task-level paired differences in answer overlap and marked usage. These diagnostics are operational proxies rather than independently validated measures of memory quality.

\begin{table*}[!t]
\centering
\caption{Memory retrieval and application behavior during task execution.}
\label{tab:rq4-memory-quality}
\scriptsize
\begin{threeparttable}
\renewcommand{\arraystretch}{1.12}
\setlength{\tabcolsep}{3pt}

\begin{tabular}{
L{0.105\textwidth}
P{0.070\textwidth}
P{0.095\textwidth}
P{0.075\textwidth}
P{0.095\textwidth}
P{0.100\textwidth}
P{0.095\textwidth}
P{0.075\textwidth}
P{0.095\textwidth}
P{0.080\textwidth}
}
\toprule
\rowcolor{gray!15}
\textbf{Memory system}
& \textbf{Diagnostic runs}
& \textbf{Retrieved per run}
& \textbf{Written per run}
& \textbf{Answer overlap}
& \textbf{Non-contradiction}
& \textbf{Marked usage}
& \textbf{Failed retrieval}
& \textbf{Incorrect-use label}
& \textbf{Mean progress} \\
\midrule

Mem0
& 333
& 11.408 (4.319)
& 6.000 (0.000)
& \textbf{0.111 (0.036)}
& 0.998 (0.017)
& \textbf{0.842 (0.138)}
& 27
& 0
& \textbf{0.316} \\

Hindsight
& 360
& 14.536 (1.184)
& 6.000 (0.000)
& 0.098 (0.031)
& 1.000 (0.006)
& 0.712 (0.151)
& 0
& 1
& 0.311 \\

Acontext
& 360
& 14.675 (2.470)
& 6.000 (0.000)
& 0.039 (0.014)
& 1.000 (0.000)
& 0.148 (0.138)
& 0
& 327
& 0.261 \\

\bottomrule
\end{tabular}

\begin{tablenotes}
\scriptsize
\item Note. Values in parentheses are run-level standard deviations. Counts and mean progress use all 360 runs per system. Mem0 diagnostic means use the 333 runs with successful retrieval because its 27 failed-retrieval runs have undefined diagnostic values. The incorrect-use count reports assignments of the harness-derived \texttt{incorrect\_use\_of\_retrieved\_memory} label, not independently verified errors. Written-memory count is fixed at six per run by design.
\end{tablenotes}

\end{threeparttable}
\end{table*}

\begin{table}[!t]
\centering
\caption{Task-level paired differences in answer overlap and marked usage.}
\label{tab:rq4-quality-contrasts}
\scriptsize
\begin{threeparttable}
\renewcommand{\arraystretch}{1.08}
\setlength{\tabcolsep}{3pt}

\begin{tabular}{
@{}
L{0.18\columnwidth}
L{0.29\columnwidth}
P{0.06\columnwidth}
P{0.12\columnwidth}
P{0.11\columnwidth}
P{0.06\columnwidth}
@{}
}
\toprule
\rowcolor{gray!15}
\textbf{Metric}
& \textbf{Contrast}
& \textbf{$n$}
& \textbf{Mean diff.}
& \textbf{Holm $p$}
& \textbf{$d_z$} \\
\midrule

\multirow[c]{3}{0.18\columnwidth}{\raggedright Marked usage}
& Mem0 vs Hindsight
& 40
& \textbf{0.123}
& \textbf{$<0.001$}
& 1.55 \\

& Mem0 vs Acontext
& 40
& \textbf{0.682}
& \textbf{$<0.001$}
& 8.14 \\

& Hindsight vs Acontext
& 40
& \textbf{0.564}
& \textbf{$<0.001$}
& 7.81 \\

\midrule

\multirow[c]{3}{0.18\columnwidth}{\raggedright Answer overlap}
& Mem0 vs Hindsight
& 40
& \textbf{0.015}
& \textbf{$<0.001$}
& 1.07 \\

& Mem0 vs Acontext
& 40
& \textbf{0.072}
& \textbf{$<0.001$}
& 5.46 \\

& Hindsight vs Acontext
& 40
& \textbf{0.059}
& \textbf{$<0.001$}
& 5.53 \\

\bottomrule
\end{tabular}

\begin{tablenotes}
\scriptsize
\item Note. For each contrast, differences are first computed within available matched LLM--task--execution cells and then averaged within each task, producing 40 task-level differences. Mean difference is the first system minus the second. Two-sided paired $t$-tests are Holm-corrected across the six contrasts. Effect sizes are paired-difference $d_z$ values computed from the task-level differences.
\end{tablenotes}

\end{threeparttable}
\end{table}

{\faCheckCircle} {\textbf{Detected contradictions are rare}:}
Mean non-contradiction is 0.998 for Mem0 and approximately 1.000 for Hindsight and Acontext. The implemented heuristic therefore rarely detects conflict with the task fact ledger. These values do not establish that the retrieved records are correct, relevant, or optimal.

{\faChartLine} {\textbf{Mem0 has the highest measured answer overlap and marked usage}:}
Among runs with defined diagnostics, Mem0 obtains the highest answer overlap (0.111) and marked usage (0.842), together with the highest mean \textit{progress\_score} among the memory-enabled systems (0.316). Its absolute answer overlap nevertheless remains low. Moreover, Mem0's diagnostic means exclude 27 failed-retrieval runs, while its retrieval count and progress include all 360 runs. Its diagnostic advantage is therefore conditional on successful retrieval.

{\faLayerGroup} {\textbf{Answer overlap and marked usage distinguish the systems}:}
All six task-level contrasts remain significant after Holm correction. Mem0 exceeds Hindsight and Acontext on both diagnostics, while Hindsight exceeds Acontext. Excluding the fallback-affected Hindsight and GPT-5.5 condition changes Hindsight's mean answer overlap from 0.098 to 0.089 and its marked usage from 0.712 to 0.751, but does not change this ordering.

{\faExclamationTriangle} {\textbf{Retrieval volume and failure labels reveal different bottlenecks}:}
Acontext retrieves slightly more records per run than Hindsight and substantially more than Mem0, but has the lowest answer overlap (0.039) and marked usage (0.148). Mem0 is primarily assigned cross-step inconsistency (283 runs), followed by model-reasoning error (45), failed retrieval (27), and no failure label (5). Hindsight has 318 cross-step inconsistency labels, 38 model-reasoning errors, one incorrect-use label, and three runs without a failure label. Acontext has 327 incorrect-use labels, 30 cross-step inconsistency labels, and three model-reasoning-error labels. These labels reflect the harness's ordered classification logic rather than independent manual verification.

{\faLink} {\textbf{Some diagnostics are associated with progress}:}
Using complete-case task-demeaned observations, inference adjusted for the 40 task effects, and Holm correction across 11 estimable correlations, Mem0 progress is associated with answer overlap ($r=0.345$, Holm $p<0.001$) and retrieved-record count ($r=0.237$, Holm $p<0.001$). Acontext progress is associated with marked usage ($r=0.232$, Holm $p<0.001$) and retrieved-record count ($r=0.162$, Holm $p=0.030$). The Hindsight marked-usage association does not remain significant ($r=0.139$, Holm $p=0.089$). These modest associations do not establish that the diagnostics caused the observed progress differences.

\begin{rqanswerbox}
\noindent\rqanswericon\hspace{0.35em}\textbf{Answer to RQ4:}
Retrieval volume alone does not indicate effective memory application. Conditional on successful retrieval, Mem0 achieves the highest measured answer overlap and marked usage, while Hindsight shows intermediate values. Acontext retrieves slightly more records than Hindsight but has substantially lower answer overlap and marked usage. Non-contradiction remains near the ceiling for all three systems and provides little discrimination. Overall, retrieval reliability, answer overlap, and marked usage capture distinct aspects of memory behavior, none of which alone explains task progress.
\end{rqanswerbox}

\section{Discussion}
\label{sec}

\subsection{Memory Improves Progress, but Does Not Ensure Task Completion}
\label{sec}

The central finding of this study is that memory improves partial progress in interdependent tasks, but does not ensure reliable task completion. In RQ1, mean \textit{progress\_score} increases from 0.238 for the stateless baseline to 0.316 for Mem0, 0.311 for Hindsight, and 0.261 for Acontext. Mem0 and Hindsight remain significant under the task-level paired analysis, whereas the smaller Acontext gain does not remain significant after task-level correction.

The results define the boundary of this benefit. Full-task success remains near the floor: Pass@1 is below 1.5\% for all memory-enabled systems, while Pass@3 reaches only 2.5\% for Mem0 and Hindsight. The evidence for partial-progress improvement is therefore stronger than the evidence for task completion. This distinction matters because progress-oriented and completion-oriented measures support different conclusions about practical readiness. Final success alone can conceal incremental gains, whereas progress alone can overstate end-to-end task-solving ability.

This finding is consistent with prior work showing that structured reasoning, feedback, and memory can support multi-step LLM behavior while long-horizon task completion remains difficult. ReAct demonstrates the value of combining reasoning and action~\cite{yao2023react}, while Reflexion shows that feedback and episodic memory can inform subsequent decisions~\cite{shinn2023reflexion}. Surveys also identify memory as a mechanism for sustaining extended interactions~\cite{zhang2025memorysurvey}. Our results qualify these expectations: in the evaluated MemoryArena-based setting, memory supports partial progress but does not resolve the full-task completion bottleneck.

Memory-enabled systems should therefore be evaluated using both progress and completion measures. The evaluated systems are better understood as progress-supporting configurations than as reliable end-to-end task solvers. Their use in consequential workflows would still require validation, fallback mechanisms, and human oversight.

\subsection{Useful Memory Depends on Writing, Retrieval, and Application}
\label{sec}

The RQ2 and RQ4 results indicate that useful memory depends on an interaction chain: information must be written, retrieved when needed, and applied in the generated response. In the primary analysis, Mem0 shows Holm-corrected gains at subtasks 1, 3, and 5, whereas Hindsight shows gains at subtasks 1 and 3. The Mem0 subtask-5 result becomes borderline under the task-level sensitivity analysis, and the Hindsight gains are sensitive to the fallback-affected GPT-5.5 condition. The position-specific pattern is consistent with memory becoming useful after task information has been written and made available to later subtasks, but it does not show that particular retrieved records caused the gains.

The retrieval trajectories support this interpretation while also showing its limits. Mem0 retrieves no records at subtask 0, but its mean retrieval count reaches 2.775 by subtask 5. Hindsight reaches the retrieval limit of 3.000 records from subtask 3 onward. Acontext retrieves two server-supplied skill descriptions at subtask 0 and remains comparatively stable thereafter, but it shows no corrected position-specific performance advantage. Access to stored information is therefore insufficient unless the retrieved content supports the current response.

Taken together, the results expose several possible failure points. A system may fail to preserve useful information, retrieve it at the appropriate step, return aligned content, or enable the LLM to incorporate that content effectively. This interpretation aligns with work that treats memory as an active process rather than a passive store. Generative Agents combine stored experiences, reflection, and retrieval to support later behavior~\cite{park2023generativeagents}, Reflexion stores feedback for subsequent decisions~\cite{shinn2023reflexion}, and MemGPT separates memory management from immediate context use~\cite{packer2023memgpt}. Our findings show that enabling a memory interface alone does not ensure measurable benefit.

\subsection{Memory Effectiveness Depends on the LLM}
\label{sec}

RQ3 shows that memory effectiveness depends on the LLM with which the memory system is combined. Claude Opus 4.8 achieves the highest overall mean \textit{progress\_score}, whereas GPT-5.5 exhibits the largest gains from Mem0 and Hindsight. For GPT-5.5, Mem0 improves mean progress by 0.165 and Hindsight by 0.125 relative to the matched stateless condition. The Hindsight result must be interpreted in light of the adapter fallback used in this configuration. For Claude Opus 4.8, the corresponding gains are smaller at 0.072 and 0.069, but both remain significant after correction. For Gemini 3.5 Flash, the observed differences are small and not statistically reliable.

These results distinguish absolute LLM performance from responsiveness to memory. The strongest LLM in absolute performance is not necessarily the LLM that benefits most from memory augmentation. The significant memory-system and LLM interaction ($\eta_p^2=0.097$) is driven primarily by the GPT-5.5 results. The interaction remains significant when Hindsight is excluded ($p<0.001$), indicating that the adapter fallback does not fully explain it. In contrast, the interaction does not remain significant when GPT-5.5 is excluded ($p=0.232$).

This result is consistent with prior work showing that the value of reasoning, feedback, and memory support depends on the underlying model's ability to use the provided information. ReAct requires the model to integrate reasoning and action~\cite{yao2023react}, while Reflexion requires it to interpret and reuse feedback~\cite{shinn2023reflexion}. Our results provide corresponding evidence for memory augmentation: the same memory system can produce gains with one LLM and limited gains with another. Memory systems should therefore be evaluated with the intended LLM rather than selected independently from it.

\subsection{Retrieval Volume Alone Does Not Indicate Memory Quality}
\label{sec}

RQ4 shows that retrieval volume alone is not a sufficient cross-system indicator of effective memory behavior. Acontext retrieves the most records per run on average (14.675), followed by Hindsight (14.536) and Mem0 (11.408). Its two server-supplied skill descriptions contribute to the Acontext retrieval count at subtask 0. Nevertheless, Mem0 achieves the highest mean progress, answer overlap, and marked usage, whereas Acontext exhibits the lowest answer overlap and marked usage. Retrieving more records therefore does not necessarily correspond to stronger alignment with the generated answer or better task performance.

The operational diagnostics clarify this pattern but require careful interpretation. Mean answer overlap is 0.111 for Mem0, 0.098 for Hindsight, and 0.039 for Acontext. Because this indicator is based primarily on token-set Jaccard overlap between retrieved records and the generated answer, it is not a direct measure of semantic relevance to the question or expected answer. Similarly, marked usage indicates that a record was classified as used through the harness procedure. It does not establish that the record was relevant, necessary, or sufficient for completing the subtask. Mem0's diagnostic means also exclude its 27 failed-retrieval runs and therefore characterize behavior conditional on successful retrieval.

Retrieved records are almost never flagged as conflicting with the task fact ledger. The primary observed differences therefore concern answer overlap and marked usage rather than detected contradiction. However, non-contradiction is derived using token overlap and a negation heuristic and should not be interpreted as independent verification of factual correctness. A system can retrieve many non-conflicting records while providing limited benefit if those records are weakly connected to the current response.

The within-task correlations further show that no single diagnostic serves as a universal proxy for memory effectiveness. For Mem0, answer overlap ($r=0.345$) and retrieval count ($r=0.237$) are positively associated with task progress after correction. For Acontext, marked usage ($r=0.232$) and retrieval count ($r=0.162$) show smaller positive associations. The corresponding Hindsight marked-usage association does not remain significant after correction. These modest, system-specific relationships reinforce the need to interpret the indicators together and do not establish causal effects.

The failure-category distributions also indicate different observable bottlenecks. Mem0 includes retrieval-failure cases, whereas Acontext is predominantly assigned the incorrect-use category. These categories result from the harness's ordered classification logic and should be interpreted as diagnostic labels rather than independently verified causes. The Acontext findings should also be interpreted under the within-task reset protocol. The study does not establish whether broader histories or accumulated skills would change its behavior.

These findings are consistent with research on retrieval-augmented generation and long-context use. Retrieval-augmented generation depends on retrieving evidence that can support generation~\cite{lewis2020rag}, while long-context studies show that placing relevant information in context does not guarantee its reliable use~\cite{liu2024lostinthemiddle}. Long-term conversational-memory benchmarks similarly show that maintaining and applying information across extended interactions remains difficult~\cite{maharana2024locomo}. Our results extend this observation to implemented memory-system configurations: retrieval activity alone provides limited evidence about memory effectiveness.

\subsection{Implications for Research and Practice}
\label{sec}

Taken together, the four research questions show that memory-enabled systems require diagnostic evaluation beyond aggregate outcomes. Memory can improve partial progress without ensuring completion, its effects emerge at particular task positions, its effectiveness varies across LLMs, and retrieval volume does not necessarily reflect effective application. Memory should therefore be evaluated as an end-to-end systems capability involving writing, retrieval, selection, application, and interaction with the underlying LLM.

For software engineering research, these findings identify evaluation concerns relevant to AI-supported software evolution and engineering workflows. Such workflows often contain dependent steps: a change request may introduce new constraints, an impact analysis may produce intermediate findings, a refactoring may depend on previous design decisions, and documentation updates may need to remain consistent with earlier artifact changes. The present benchmark results do not demonstrate memory effectiveness on these specific activities. Instead, they show why future evaluations should examine whether earlier constraints are preserved, appropriate information is retrieved, and that information is consistently applied during later steps.

For researchers, future evaluations should distinguish partial progress from full-task success, report position-level behavior, analyze memory gains separately for each LLM, and examine operational indicators together with task outcomes. Evaluations should also preserve record-level lineage from writing through retrieval and application. Such provenance would enable stronger analysis of whether a retrieved record contributes to a subsequent decision and reduce reliance on indirect diagnostic measures.

For practitioners, adding memory should be treated as a configuration and measurement decision. Teams should evaluate the complete LLM--memory configuration using representative task sequences and realistic interaction constraints. They should inspect retrieval failures, answer overlap, marked usage, and downstream outcomes rather than relying solely on retrieval counts or aggregate success. The evaluated systems also reveal different priorities for investigation. Mem0's retrieval failures motivate attention to retrieval reliability, whereas Acontext's high retrieval volume and low marked usage motivate attention to selection and response conditioning. These implications concern the evaluated configurations and should be validated in the intended deployment setting.
\section{Threats to Validity}
\label{sec:threats}

\textbf{Construct validity.}
Answer overlap, non-contradiction, and marked usage are automatically derived proxies rather than independently human-validated assessments, and we therefore treat them as diagnostic indicators rather than measures of memory quality.
For Mem0, these indicators are calculated only when retrieval succeeds, and failures are reported separately.
Low answer overlap indicates weak token-level alignment with the generated answer rather than low relevance to the question, non-contradiction indicates absence of detected conflict with the task fact ledger rather than perfect memory quality, and marked usage indicates records classified as used through \textit{used\_memory\_ids} or an overlap-based or LLM-based fallback rather than proof that retrieved content was relevant, necessary, or sufficient.
Task outcomes also depend on the fixed completion threshold of 0.6, and a different cutoff could change \textit{progress\_score} and strict success.
Another construct-validity threat concerns the judge.
Responses are scored by GPT-4o-mini, which comes from the same provider as GPT-5.5 and serves as the internal processing LLM for all memory systems.
This may introduce provider-associated or style-preference bias.
Using the same judge standardizes scoring across configurations but does not eliminate differential bias across LLMs.
Future work should repeat scoring with independent judges or human assessment to determine whether the rankings remain stable.

\textbf{Internal validity.}
The results may be influenced by choices in the evaluation pipeline, including the prompt template, memory-write policy, retrieval configuration, scoring rubric, and system-specific formatting.
The harness caps retrieval at three records per subtask, so systems that repeatedly reach the cap may be truncated by the evaluation setup.
Observed retrieval volume therefore reflects both memory-system behavior and the fixed retrieval limit, and Acontext's volume includes two server-supplied skill descriptions returned before any benchmark interaction is written.
The retrieval query is also fixed: the raw current subtask question is submitted as the query.
Query rewriting, constraint-directed retrieval, or explicit memory queries could change retrieval quality and downstream usage.
The write policy is fixed as well.
After each subtask, the harness writes one record containing the current question and final answer.
This protocol does not store full reasoning traces, rejected alternatives, tool traces, or all constraints considered during generation.
The comparison should therefore be interpreted as an evaluation of the memory systems under a shared forced-write protocol rather than a complete assessment of each system's native memory-management strategy.
In addition, all Hindsight and GPT-5.5 observations invoked the adapter's local fallback at least once, and several Hindsight findings are sensitive to excluding this condition.
The event logs do not preserve execution identifiers, so RQ2 averages pooled observations rather than reconstructing execution-specific trajectories, and incomplete event traces were replaced with traces generated under the same protocol.
Moreover, the study does not trace the complete lineage of individual facts from writing through retrieval to downstream application.
The RQ2 and RQ4 findings therefore identify patterns consistent with within-task reuse, but do not establish that a particular retrieved memory caused an observed performance improvement.

\textbf{External validity.}
The evaluation covers one MemoryArena domain, four conditions, three LLMs, 40 tasks, and three repeated executions per condition.
The findings may not generalize to other domains, longer interaction horizons, alternative execution pipelines, alternative memory systems, or updated model versions.
Because memory is reset before each task and repeated execution, the study does not examine cross-task transfer, accumulated reusable skills, open-ended multi-session memory, persistent organizational memory, or privacy-sensitive settings.
This is particularly relevant for Acontext, whose skill-learning design may benefit from reuse across broader histories than those evaluated here.
The stateless baseline receives neither external memory nor previous conversation history, and we do not compare against a long-context baseline containing prior interaction history.
The study therefore does not determine whether external memory provides benefits over retaining the full interaction history in context.
The evaluated tasks provide interdependent sequences in which later steps depend on earlier choices, constraints, and intermediate information.
These properties are relevant to memory-enabled LLM systems for AI-supported software evolution and engineering workflows, but future work should evaluate the same systems on software-specific artifacts such as requirements, issue reports, code changes, architecture decisions, and technical documentation.
Real software evolution workflows may also involve longer memory horizons, richer artifacts such as issue states, code review comments, tool outputs, and design rationale, and decisions that are revised or contradicted over time.

\textbf{Conclusion validity.}
We apply Holm correction within the specified comparison families and report effect sizes alongside statistical significance.
However, the observed effects are not uniform: the memory-system and LLM interaction is statistically significant and driven primarily by GPT-5.5, while Acontext exhibits only weak and analysis-dependent gains over stateless.
The task-blocked repeated-measures ANOVA summarizes the overall effects of memory system and LLM, but it averages repeated executions within each cell.
Because a mixed-effects model was not fitted, the interaction should be interpreted as evidence of configuration-dependent variation rather than a precise estimate of all variance sources.
The execution identifiers are not passed as sampling seeds to the LLM APIs, so variation across repeated executions also reflects uncontrolled sampling behavior.
For RQ2, the primary analysis contains three LLM observations for each task; the task-level sensitivity analysis reduces this dependence and preserves the principal subtask-1 and subtask-3 findings.
The near-floor Pass@$k$ values indicate a floor effect for strict completion, limiting our ability to distinguish whether memory has little effect on completion or whether the benchmark setting is too difficult for reliable full-task success under all configurations.
Accordingly, the evidence for partial-progress improvement is stronger than the evidence for differences in reliable task completion.

\textbf{Reproducibility and deployment boundary.}
The study relies on hosted LLMs and evolving memory-system implementations, and exact replication may therefore be affected by changes in these services.
The harness is custom and is not the official MemoryArena runner. The experiment code, including the custom harness and analysis scripts, is publicly available in an anonymous replication repository~\cite{memory-replication}. The study dataset, including task selections, execution logs, retrieval logs, and scoring outputs, is archived separately~\cite{memory-dataset}. The study does not evaluate operational concerns such as cost, latency, privacy, memory poisoning, stale memories, user trust, or human acceptance. The results therefore characterize memory systems within a MemoryArena-based evaluation and do not establish readiness for unsupervised use in high-impact settings.
\section{Conclusion}
\label{sec:conclusion}

This study evaluated external memory across 40 MemoryArena-based tasks, three LLMs, and three repeated executions, yielding 1,440 task-level observations. Mem0 and Hindsight improved partial progress, whereas Acontext's smaller gain was not robust under task-level analysis and full-task success remained near the floor. Effects varied across subtasks and LLMs, and greater retrieval volume did not consistently correspond to higher answer overlap, marked usage, or progress. The next phase of the industrial collaboration will evaluate memory-enabled LLM systems in a planned production deployment, focusing on software-intensive workflows, longer interaction histories, effectiveness, reliability, latency, privacy, and monitoring.

\section*{Data Availability}
The experiment code~\cite{memory-replication} and the study dataset~\cite{memory-dataset} are publicly available.

\bibliographystyle{ACM-Reference-Format}
\bibliography{references}

\end{document}